# Amplified Urban Climate Extremes from Global Warming-Urbanization Synergy: A Physics-Informed Intelligence Framework

Qiuxia Wu*, Yaqiang Wang, Huabing Ke

Institute of Artificial Intelligence for Meteorology, Chinese Academy of Meteorological Sciences, Beijing 100081, People's Republic of China

*Corresponding Author: Qiuxia Wu, wuqx@cma.cn

## Abstract

The nonlinear synergy between global warming and urbanization is amplifying extreme climate risks in cities worldwide. While observations and simulations confirm these compounding effects, two fundamental bottlenecks impede predictive understanding: (1) fragmented, case-specific perspectives that hinder the discovery of universal mechanisms, and (2) a methodological divide between computationally prohibitive high-resolution models and AI-based tools that lack physical interpretability at urban scales. This article advocates for a paradigm shift toward the deep integration of physical principles with data intelligence. To this end, we propose a transformative "Classification-Mechanism-Inference" (CMI) framework. *Classification* involves establishing a global urban "climate-morphology-development" typology to enable systematic comparison beyond isolated case studies. *Mechanism* advocates for physics-informed machine learning (PIML) as the core engine to develop efficient, physics-constrained surrogate models for uncovering nonlinear interactions. *Inference* leverages these models for high-throughput, tailored risk projection to directly inform context-specific adaptation planning. The CMI framework aims to bridge the cognitive and methodological gaps, thereby advancing urban climate science from phenomenological description towards mechanistic, predictive, and decision-relevant science, which is crucial for building climate-resilient cities globally.

## Introduction

The IPCC's Sixth Assessment Report finds that the 2011–2020 global mean temperature was 1.09 °C above the 1850–1900 average, and projects that the 1.5 °C warming level is likely to be exceeded within the next two decades (IPCC, 2021). This trend has driven substantial increases in the frequency, intensity, and duration of extreme heatwaves, heavy precipitation, and compound events, resulting in a cascading amplification of systemic climate risks (Trenberth, 2008; Diffenbaugh et al., 2017; Raymond et al., 2020). While climate change is a global phenomenon, its impacts are highly heterogeneous, with urban areas facing disproportionately severe risks. The interaction between global-scale climate forcing and local urbanization—primarily through land-cover modification and anthropogenic heat release, and most visibly as the urban heat island (UHI) effect—has long been recognized (Alcoforado and Andrade, 2008). The precise mechanisms driving this interaction, however, particularly the nature of their complex, nonlinear synergy, have remained unclear. Recent research provides stronger mechanistic insight and evidence that global warming and urbanization are not independent drivers. Rather, they engage in mutual amplification through multiple physical feedbacks, producing a combined impact greater than the sum of their individual effects. This synergy significantly exacerbates specific hazards, particularly the intensity and frequency of extreme heat in cities (Creutzig et al., 2025; Liao et al., 2025; Khan et al., 2025). Studies of Chinese cities, for example, show that greenhouse gas emissions and urban expansion have combined to increase compound extreme heat events, with the attributable contributions of each driver being separable and quantifiable (Wang J. et al., 2021). This amplification extends beyond temperature. The spatial configuration of urban areas—including compact development—also shapes extreme rainfall by altering local thermal and dynamic conditions (Yang et al., 2024). As a result, urban areas serve not only as hotspots that amplify climate risks but also as increasingly vulnerable systems (Sarangi et al., 2021; Zhou et al., 2022). Emerging research indicates that this synergy between global warming and urbanization is becoming a key driver of future urban environments and human well-being (Sethi and Vinoj, 2024; Luo et al., 2024). Consequently, accurately disentangling the physical mechanisms and spatial heterogeneity of this synergy constitutes a core scientific challenge at the intersection of urban climatology, climate adaptation, and sustainability science. Meeting this challenge calls for new research paradigms that fuse physical principles with data-driven intelligence—an urgent necessity for advancing climate-resilient urban development.

To develop such a paradigm, a systematic breakthrough is needed, beginning with a representative field where mechanisms are clear and evidence is robust. Extreme high temperatures—heatwaves—represent a key focal point for analyzing the non-linear synergy between global warming and urbanization. Their physical drivers share origins with the urban heat island effect, and they benefit from the most extensive foundation of observational and attribution studies. Therefore, this paper takes extreme high temperatures as a starting point. We systematically synthesize progress on their synergistic amplification and methodological bottlenecks, and propose a new "Classification–Mechanism–Inference" (CMI) research paradigm. This framework aims to provide a transferable foundation and methodological pathway for understanding synergies in other extreme climate events and for developing intelligent analysis tools. This paper first reviews the observed characteristics, multidimensional impacts, and complex adaptation challenges associated with the synergistic amplification of extreme heat by global warming and urbanization (Section 1). It then analyzes the methodological evolution—and current bottlenecks—from high-resolution simulations to emerging artificial intelligence paradigms (Section 2), distilling the key challenges in achieving both a systematic understanding and methodological integration (Section 3). To overcome these challenges, we propose and detail a new "Classification–Mechanism–Inference" (CMI) paradigm, designed to deeply integrate physical mechanisms with data intelligence, and present its core framework (Section 4). Through this synthesis, the paper aims to promote a paradigm shift in this research field and provide a clear pathway for deepening mechanistic understanding and developing intelligent tools in the future.

# 1. Synergistic Amplification: Observational Evidence, Impacts, and Adaptation

## 1.1. Intensification of Extreme Heat and Its Modulation by Urbanization

Global warming is the primary driver behind the intensification of extreme climate events (IPCC, 2021, SPM B.2). Climate model projections consistently show that heatwaves will become more intense, frequent, and prolonged in the future (Meehl and Tebaldi, 2004). The occurrence of extreme heatwaves results from a close synergy between large-scale anthropogenic forcing—primarily greenhouse gas-induced warming—and rapid urbanization. Global research consistently reveals this multi-scale amplification. In North America, for instance, high-pressure blocking systems such as "heat domes" are key circulation drivers of extreme heat (Zhang et al., 2023). Under a warming climate, the intensity of these events is further enhanced by soil moisture-atmosphere feedbacks. Concurrently, the persistent, large-scale urban heat island effect provides a continuous thermal baseline. A study of 50 major U.S. cities shows an average surface UHI intensity of 2.88°C, with most cities experiencing a decades-long intensification trend (Xian et al., 2023). When synoptic-scale heatwaves occur, they interact with these intensified local urban heat islands, compounding thermal stress in metropolitan areas through multi-scale feedbacks (Chen et al., 2023). In Europe, the synergy between the North Atlantic Oscillation and atmospheric blocking provides a key large-scale circulation background for persistent heatwaves (Vautard et al., 2023; Li et al., 2020). At the urban scale, studies show that such heatwaves can markedly intensify nocturnal urban heat islands. This interaction exacerbates thermal stress and health risks in urban cores (Possega et al., 2022; Founda and Santamouris, 2017). In Africa, large-scale attribution studies identify anthropogenic warming as the primary driver of recent heatwave trends (Bobde et al., 2025), while observational research shows that urbanization further intensifies nocturnal heatwaves in cities through the urban heat island effect (Igun et al., 2023). In Oceania, exemplified by Australia, urban overheating results from a combination of local heat islands and secondary circulations such as sea breezes (Yenneti et al., 2020). Urbanization-induced microclimatic differences cause the number of hot days (daily maximum temperature ≥35 °C) recorded within cities to substantially exceed suburban station records. For example, in some urban zones, micro-stations have recorded more than four times as many hot days as nearby official weather stations (32 days versus 7 days), significantly amplifying population heat exposure risks (Pfautsch et al., 2025). In Eurasia, the variability of the East Asian monsoon system sets an important climatic backdrop for regional extremes (Huang and Du, 2010). Recent studies indicate that under global warming, dry – wet variability in the monsoon region has intensified, leading to more frequent extreme weather and climate events (Chen et al., 2025). Against this background, model simulations further suggest a significant increase in the frequency of record-breaking extreme heat (Nath et al., 2024). Attribution studies quantify that global warming and urbanization contribute approximately 72.9% and 21.9%, respectively, to the intensification of human-perceived heatwaves in China (Kong et al., 2020). In rapidly developing urban regions such as the Beijing-Tianjin-Hebei agglomeration, urbanization alone can explain 27% to 68% of the rise in extreme temperature indices in core cities (Wang Y. et al., 2021). Results from multiple attribution approaches converge to show that urbanization is a key driver of extreme heat intensification, yet quantitative estimates remain inconsistent across studies.

The synergy between the aforementioned large-scale circulation patterns and local urbanization forms a key framework for heatwave intensification. Urbanization not only amplifies the urban heat island by altering surface energy and water balances, but more critically, it modulates the spatiotemporal characteristics and impacts of extreme heat events. This modulating role is especially evident in rapidly developing megacities, such as Shanghai on China’s eastern coast.

Taking Shanghai as a case study, long-term near-surface observations from 10 national meteorological stations across the municipality—spanning 70 – 80 years (Fig. 1a)— clearly illustrate the local climatic imprint of rapid urbanization. The analysis (Fig. 1b) shows that Xujiahui station, located in the highly urbanized city center, has warmed by about 3 ° C since 1951, a trend substantially steeper than the global average over the same period. In contrast, Chongming station, representing a more natural suburban setting, exhibits a much more moderate warming trend. This pronounced urban – rural warming gradient strongly indicates that rapid urbanization is a key driver of intensified local heat islands and spatially heterogeneous warming (Kalnay and Cai, 2003; Yang et al., 2011; Liao et al., 2017; Zhou et al., 2022). Attribution analyses further show that, across the Yangtze River Delta and China's major urban clusters, urbanization accounts for roughly one-sixth of the observed warming in city centers (Wang J. et al., 2021). Recent studies offer a clearer mechanistic understanding: urbanization alters the surface energy balance, intensifying the urban heat island and enhancing heat retention, which significantly increases population exposure to extreme heat and associated public health risks (Liao S. et al., 2025; Liao W. et al., 2025; Yu et al., 2026). These works delineate the specific physical pathways through which the synergistic framework described earlier operates at the urban scale.

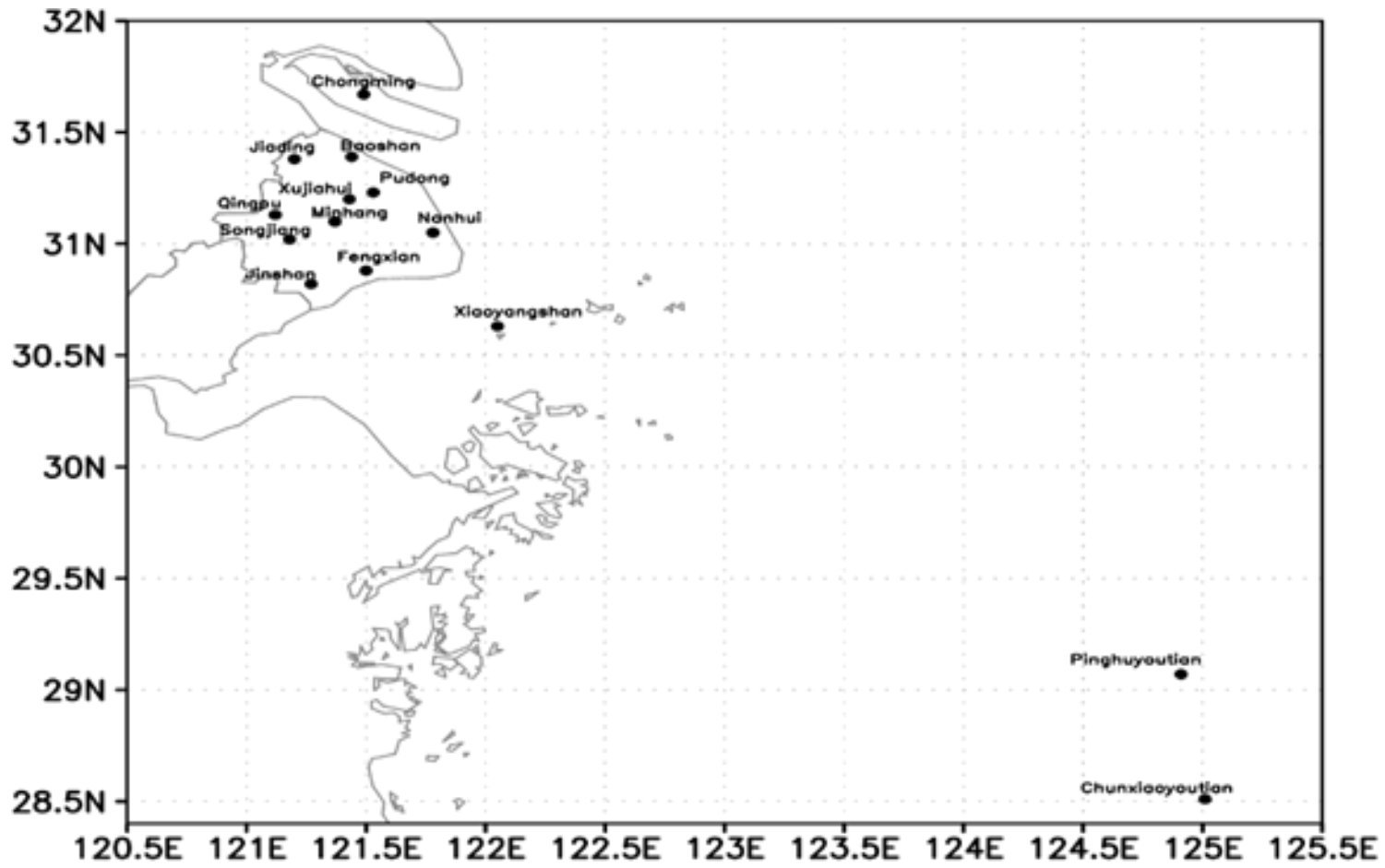

**Fig. 1a. Meteorological station network in Shanghai, China.** The ten marked stations (e.g., Xujiahui, Chongming) provide 70–80 years of observational data, with records extending from the 1950s onward, offering a long-term observational foundation for climate-change research in Shanghai.

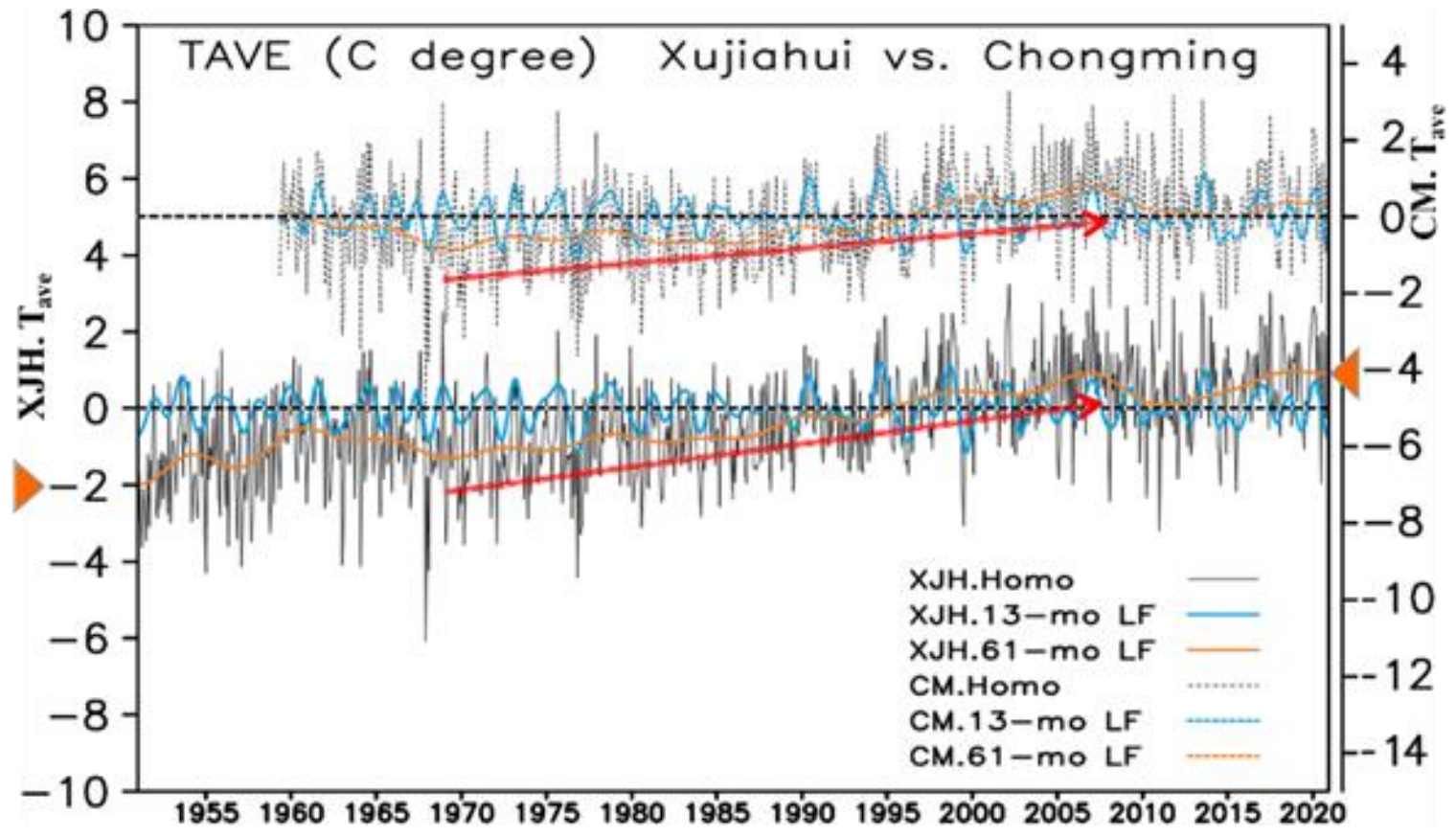

**Fig. 1b. Comparison of monthly temperature anomaly series (1951–2020) between an urban (Xujiahui, XJH, left axis) and a suburban (Chongming, CM, right axis) station in Shanghai.** The magnitude and rate of warming are substantially greater at

the urban station, highlighting the amplification of local warming by urbanization. The homogenized series are displayed as: the raw series (black line), a 13–60-month bandpass filter (blue), a 5-year low-pass filter (orange), and the linear trend (red line).

Synthesizing global observations and attribution research, extreme heat risk is fundamentally governed by a complex, multi-scale synergy of large-scale circulation anomalies (driven by global warming) and local urban processes. This synergy extends beyond the short-term amplification of the urban heat island during extreme events; it systematically reshapes the long-term urban climate background, thereby establishing megacities as a primary frontier and amplifier for climate-risk escalation.

## 1.2. Multidimensional Impacts, Resilience Challenges, and the Complexity of Adaptation

Synergistically amplified climate risks pose a severe, escalating threat to socio-ecological systems (IPCC, 2022; Raymond et al., 2020), driving a fundamental paradigm shift in the broader field of climate risk research from single-hazard analysis to the study of compound, systemic impacts (Lempert, 2021). This shift is motivated by key realizations: historical extremes like the 2003 European heatwave demonstrated how a single event triggers cascading failures across health, energy, and agriculture (de Bono et al., 2004), while global assessments show a rapid expansion in populations exposed to lethal heat stress, heightening the urgency for systemic responses (Mora et al., 2017). This broader paradigm shift is especially critical for rapidly urbanizing regions, where research details the synergy between global warming and urbanization. This interplay amplifies heatwaves (Kong et al., 2020), asymmetrically reshapes precipitation (Xiong et al., 2025), and modulates drought evolution (Huang S. Z. et al., 2022; Huang X. et al., 2022). These findings establish urban areas as the pivotal frontier for confronting compound climate risks.

Cutting-edge adaptation strategies call for coordinated, multi-scale interventions—spanning urban planning, blue-green infrastructure, and community resilience (IPCC, 2022, Chapter 6; Cook et al., 2025). These strategies operate across several interconnected fronts: enhancing urban-ecosystem resilience through the protection and restoration of biodiversity and associated ecosystem services (Elmqvist et al., 2013); developing quantitative assessment models to guide blue-green infrastructure planning (Haase et al., 2014); quantifying the cooling effects of vegetation to support nature-based solutions (NbS) (Bounoua et al., 2015); integrating socio-ecological perspectives for managing climate impacts (Hobbie and Grimm, 2020); and using downscaled climate projections and urban-morphology analyses to anticipate future heat risks and inform proactive urban design (He et al., 2022).

Emerging research complicates the picture of climate adaptation, revealing inherent trade-offs and even contradictions within strategies themselves. A landmark global study of over 3,000 cities uncovered a "dual health effect" of urban heat islands: while they exacerbate heat-related mortality in summer, they substantially reduce cold-related deaths in winter. Globally, the lives saved from cold outweigh the lives lost to heat by more than four to one (Wang S. et al., 2025). This work further demonstrates that widespread cooling interventions—like expanding green cover or using reflective surfaces—have highly variable net benefits, which depend on geography and season and can even be negative in higher-latitude cities. Compounding these complexities are serious equity concerns. For example, during extreme droughts, vegetation degradation is often most severe in marginalized communities, disproportionately intensifying their heat exposure (Yan et al., 2026). These findings underscore a critical need to evolve adaptation planning beyond one-dimensional solutions. The path forward requires integrated frameworks that explicitly account for full-seasonal impacts, navigate multi-objective trade-offs, and center social justice.

Precise, spatially explicit projections of future risk are therefore foundational to designing effective adaptation strategies, which must be integrated into coordinated actions across global, regional, and local scales. The complexity of the challenge limits the effectiveness of standardized, "one-size-fits-all" engineering approaches. Consequently, Nature-based Solutions (NbS)—defined as the protection, management, and restoration of ecosystems to address societal challenges while simultaneously enhancing climate resilience—have gained prominence as context-sensitive strategies offering multiple co-benefits and inherent adaptive capacity. This framework, now standardized and promoted by major international bodies like the International Union for Conservation of Nature (IUCN) and the United Nations Environment Assembly (UNEA) (Cohen-Shacham et al., 2016; IUCN, 2020; UNEA, 2022), directly supports the Paris Agreement's Global Goal on Adaptation (GGA) in building resilience and reducing vulnerability (UNFCCC, 2015). In response, many nations are mainstreaming such principles into national policy. China's National Climate Change Adaptation Strategy 2035, for example, which seeks to build a climate-resilient society through proactive, systemic, and synergistic action, aligns closely with this international paradigm (Ministry of Ecology and Environment et al., 2022).

Future adaptation requires cross-sectoral collaboration and solutions like NbS that deliver multiple benefits. However, the effectiveness and equity of NbS are contingent on reliable supporting systems—such as water security for urban greenery during droughts (Yan et al., 2026). Furthermore, adaptation measures can produce unintended systemic feedback; for instance, risk-based insurance may inadvertently exacerbate vulnerability through market withdrawal and disinvestment (Collier and Kirsht, 2026). Ultimately, mobilizing an effective societal response to climate risk hinges on planning grounded in a complex-systems perspective, supported by science, broad capacity building, and inclusive public engagement.

# 2. From Attribution to Projection: Methodological Evolution and Its Key Bottlenecks

## 2.1. From Observations to High-Resolution Simulation: The Legacy and Limits of Numerical Modeling

The study of urban climate is undergoing a paradigm shift: from merely describing phenomena toward a new framework that integrates mechanistic understanding, predictive capability, and planning for resilience. Early insights were built on in-situ measurements and remote sensing (Oke, 1973; Grimmond, 2007; Park et al., 2017; Luo and Lau, 2021; Xuan et al., 2026). While satellite data provide essential spatial coverage, their quantitative accuracy is limited in complex urban environments by physical constraints, such as Urban Thermal Anisotropy (UTA) caused by three-dimensional building structures, which introduce systematic directional bias. A global study of over 5,500 cities found that UTA can create an average daytime summer land surface temperature difference of 5.1 K, potentially leading to a >40% underestimation of urban sensible heat flux and surface heat island intensity derived from satellites (Du et al., 2023). This methodological foundation has since evolved to high-resolution regional climate modeling. Such models, when integrated with urban canopy schemes (e.g., WRF-Urban), provide the capability for detailed process simulation and attribution (Argüeso et al., 2013; Moraglia et al., 2024; Liao et al., 2025). They are now indispensable for a wide range of applications, including high-resolution meteorological simulation in megacities, thermal environment assessment of urban agglomerations, and fine-scale thermal comfort studies in built-up areas (Obe et al., 2024; Li et al., 2026; Huang et al., 2025; Mancheno et al., 2024). Recent integration with standardized urban classification systems like the Local Climate Zone (LCZ) has further refined their utility. Collectively, this methodological evolution—from observation to mechanistic simulation—has fundamentally advanced our understanding of urbanization's climatic effects (Stewart and Oke, 2012; Huang et al., 2025; Gál et al., 2026), providing the foundation for a more predictive science of urban climate and resilience planning.

Nevertheless, the growing sophistication of models introduces its own set of challenges. Research shows that uncertainty can stem significantly from the specification of key physical parameters, the simplification of processes, and even the code implementation within urban parameterization schemes. This underscores an urgent need to strengthen the links between physical assumptions, their computational implementation, and targeted observational validation (Joshi et al., 2025). Compounding these scientific issues is a formidable practical challenge: prohibitive computational cost. Such high-resolution, explicit simulation of complex processes imposes a major bottleneck. It severely restricts the capacity to run the high-resolution, long-term ensemble simulations needed for future climate risk assessment. It also limits the operational feasibility of kilometer-scale, rapidly-updating numerical weather prediction for effective warnings, and hinders the extensive experimentation required for systematic parameter optimization and uncertainty quantification. Consequently, computational expense remains the primary barrier preventing high-fidelity models from transitioning to scalable, actionable deployment.

While physical models grapple with fundamental scalability limits, the data-driven AI paradigm presents a transformative alternative for Earth system science, albeit with its own unique challenges.

## 2.2. The AI-Driven Paradigm Shift and the Regional-to-Urban "Last-Mile" Downscaling Challenge

Artificial intelligence is ushering in a transformative era for Earth system science (Eyring et al., 2024a, 2024b). A leading manifestation of this shift is the rise of hybrid modeling, which fuses physical principles with AI to dramatically accelerate simulations and improve parameterizations (Kochkov et al., 2024; Bracco et al., 2025; Berk et al., 2026; Luo et al., 2026). Yet, a significant gap remains for urban applications. State-of-the-art global and regional AI weather models—including GraphCast, FourCastNet, and Pangu-Weather—are fundamentally architected for synoptic-scale circulation. This design focus leaves them unable to resolve the fine-grained physics governing the urban-climate interface, such as canopy-layer energy budgets and heterogeneous surface properties (Pathak et al., 2022; Lam et al., 2023; Bi et al., 2023; Han et al., 2024). Synthesis of the literature suggests that while AI has markedly improved forecasts for extreme weather in general, the field still grapples with significant shortcomings, including a pronounced lack of targeted studies on urban-specific hazards like heatwaves (Kim and Kim, 2025). These shortcomings are rooted in part in the intrinsic difficulty AI models have with extreme events, which are inherently data-scarce and weakly-constrained by physics (Camps-Valls et al., 2025). This limitation manifests operationally as a widespread tendency for regional AI forecasts to systematically underestimate the intensity of extremes such as heavy rainfall and typhoons (Zhang et al., 2025). Therefore, applying the outputs of these biased models directly at the city scale calls their reliability and practical utility into serious question (Nath and Palmer, 2026). Together, these issues define the central "last-mile" gap in bridging AI-based regional predictions with actionable urban climate information.

By contrast, physics-based numerical models have evolved a coherent, hybrid approach that integrates mechanistic understanding with data to address urban-scale problems. This established pathway brings into sharp focus the core limitations of current AI paradigms: 1) Physical realism has progressed from bulk to multi-layer urban canopy schemes (e.g., BEP/BEM), establishing a principled process architecture (Joshi et al., 2025)—a capability still missing in current data-driven AI models, which lack embedded physical constraints. 2) Physics-based models leverage high-resolution geographic data—integrating the LCZ framework with 100-m global products (He et al., 2023)—to represent urban heterogeneity with both granularity and standardization (Huang et al., 2025; Li et al., 2026). AI models, whereas, rely primarily on atmospheric variables and are not architected to natively ingest comparably rich, high-resolution geospatial information. In resolving key processes, physics-based models can explicitly represent critical urbanization forcings—like spatially-resolved anthropogenic heat—within the climate feedback cycle (Khan et al., 2025; Liu et al., 2023). This stands in sharp contrast to current AI models, for which the data-driven representation of such physically well-defined but observationally sparse processes remains a fundamental and unresolved limitation. These advances are paving the way for physics-based "digital urban twins" that unify fine-scale physics with diverse data streams (Gál et al., 2026). Such models can now directly inform real-world decisions, from evaluating urban heat mitigation to planning for global resilience (Ding et al., 2026). Current AI-based weather models, in stark contrast, struggle to capture such multi-scale, mechanism-driven urban processes, reflecting a fundamental mismatch between their architecture and the data requirements of heterogeneous urban systems.

To overcome the "last-mile" bottleneck, a pressing need exists for a new generation of urban climate modeling frameworks that achieve a deep synthesis of physical mechanism and data intelligence. This demands a coordinated advance on three interconnected fronts. First, at the algorithmic core, the pivotal task is the deep embedding of physical constraints. This entails advancing paradigms like Physics-informed Machine Learning (PIML) to encode fundamental laws—such as conservation principles and energy balance equations—as hard constraints directly into the neural network's architecture. This deep integration forms the essential technical foundation for ensuring physical consistency, interpretability, and reliable extrapolation. Practical implementations include, for instance, the design of physics-guided loss functions (Lopez-Gomez et al., 2023). On a parallel front, achieving intelligent fusion of multi-source, heterogeneous data demands a redesigned data-model interface. The central task is to construct pipelines that feed regional meteorological fields directly into models alongside 100-meter to street-scale urban descriptors—including 3D form, land use, and anthropogenic heat. The core challenge is to bridge the scale-semantic gap between these data types, thereby enabling a"meteorology-to-urban-response"framework for block-level, high-resolution microclimate prediction. This step directly resolves the chronic under-representation of surface heterogeneity that stems from coarse model resolution. Finally, a standardized testbed for urban extreme events must be established as the third essential front. This serves as the critical foundation for the robust development of the technologies outlined above. The field must converge on accepted standards that include test protocols, benchmark datasets, and evaluation metrics (Kim and Kim, 2025; Nath and Palmer, 2026). Only with such a framework can the predictive skill of different models be fairly and objectively compared at the urban scale—especially for out-of-sample extreme events—thereby directing the community's efforts toward solving real-world urban climate problems. The overarching vision is an urban AI climate twin that achieves functional parity with—and potentially superior efficiency to—physics-based digital twins. Such a system would be capable of intrinsically simulating core urban processes, from canopy-layer energy budgets to anthropogenic heat fluxes. Through continuous verification and co-evolution with physical models, it would ultimately mature into a new generation of reliable, high-performance computational infrastructure for urban climate risk management.

The scalability limits of physics-based models and the "last-mile" gap in AI approaches together reveal a deeper dilemma: reliably understanding and predicting the synergy between global warming and urbanization requires overcoming two fundamental, interconnected systemic barriers.

# 3. The Dual Challenge of Fragmented Knowledge and Disjointed Tools

In summary, urban climate simulation and prediction faces two deeper, interlinked challenges that extend beyond the specific limitations of current physical or AI models.

First is the lack of a systematic framework. Much of our knowledge stems from isolated regional case studies. We lack a robust, globally applicable framework to systematically attribute mechanisms and compare cities across diverse climates, development stages, and spatial forms. The absence of this framework is thrown into ever sharper relief as global case studies proliferate. Syntheses of observational data demonstrate that urbanization's amplification of the urban heat island is characterized by significant spatial heterogeneity (e.g., dependence on climate zone) and nonlinear dynamics (Li et al., 2023). More strikingly, recent research indicates that nearly half of the world's cities have experienced complex trend shifts, including a deceleration in the rate of heat island intensification (Zhan et al., 2025). These results underscore that the evolution of urban thermal environments is highly heterogeneous and path-dependent, preventing the direct extraction of universal mechanisms from isolated case studies. This disconnect creates a significant gap between rich local evidence and the pursuit of a unified global model (Creutzig et al., 2025), which currently limits robust mechanistic inference and reliable prediction for diverse city types. This systemic knowledge gap directly impedes the translation of science into action. Evaluations of global urban climate action plans, for example, consistently show major shortfalls in both the comprehensiveness of their content and the comparability of their implementation standards—a clear signal that practice lacks guidance from a robust, consensus-based scientific framework (Aboagye and Sharifi, 2024). Consequently, progress hinges on a new research paradigm that moves beyond isolated case analysis to one that explicitly integrates the dynamic, multi-scale interactions of key drivers.

Developing and validating such a framework can leverage several large, internally heterogeneous nations—China, the United States, and India—as vast 'natural laboratories.' China offers a particularly powerful case, integrating a complete climate gradient from boreal to tropical zones with a process of rapid, condensed urbanization across a contiguous geography. The unique overlay of intensive human activity, spatiotemporally compressed development, and a full climatic spectrum provides a nearly ideal, fully-equipped experimental setting. Within this single, coherent analytical frame, it becomes possible to dissect the nonlinear response of the human-environment system to multiple, concurrent stressors (Miao et al., 2020). Thus, systematic investigation of this setting to uncover dominant mechanisms provides a critical pathway for bridging the systemic knowledge gap and advancing toward a more general theory.

A second, equally formidable barrier is the lack of integrative methodology. Despite the physical clarity of high-resolution simulation and the computational agility of AI, the field still lacks a new generation of analytical tools capable of truly fusing mechanistic and data-driven approaches—tools that are both physically interpretable and efficient for scenario exploration (Luo et al., 2026). This methodological gap leaves researchers unable to rapidly and rigorously dissect the dynamic, nonlinear synergy between global warming and urbanization, or to reliably quantify how this synergy modulates compound extreme event risks—all while preserving physical consistency. Consequently, this fusion bottleneck fundamentally blocks the critical progression from phenomenological attribution to reliable prediction and scenario-based warning.

## 4. A Physics-Informed CMI Framework: Toward a Closed Research Cycle in Urban Climatology

To overcome the twin challenges of fragmented understanding and disjointed methods, we call for a paradigm shift in urban climate research—one built on a deep fusion of physical mechanism and data intelligence. This shift centers on establishing an integrated Classification-Mechanism-Inference (CMI) framework, moving the field from isolated case studies to systematic insight, and from high-cost simulation toward efficient, AI-powered scenario exploration. As illustrated in Figure 2, this framework transcends traditional isolated case studies by structuring the research process into three interconnected stages: Classification, Mechanism, and Inference.

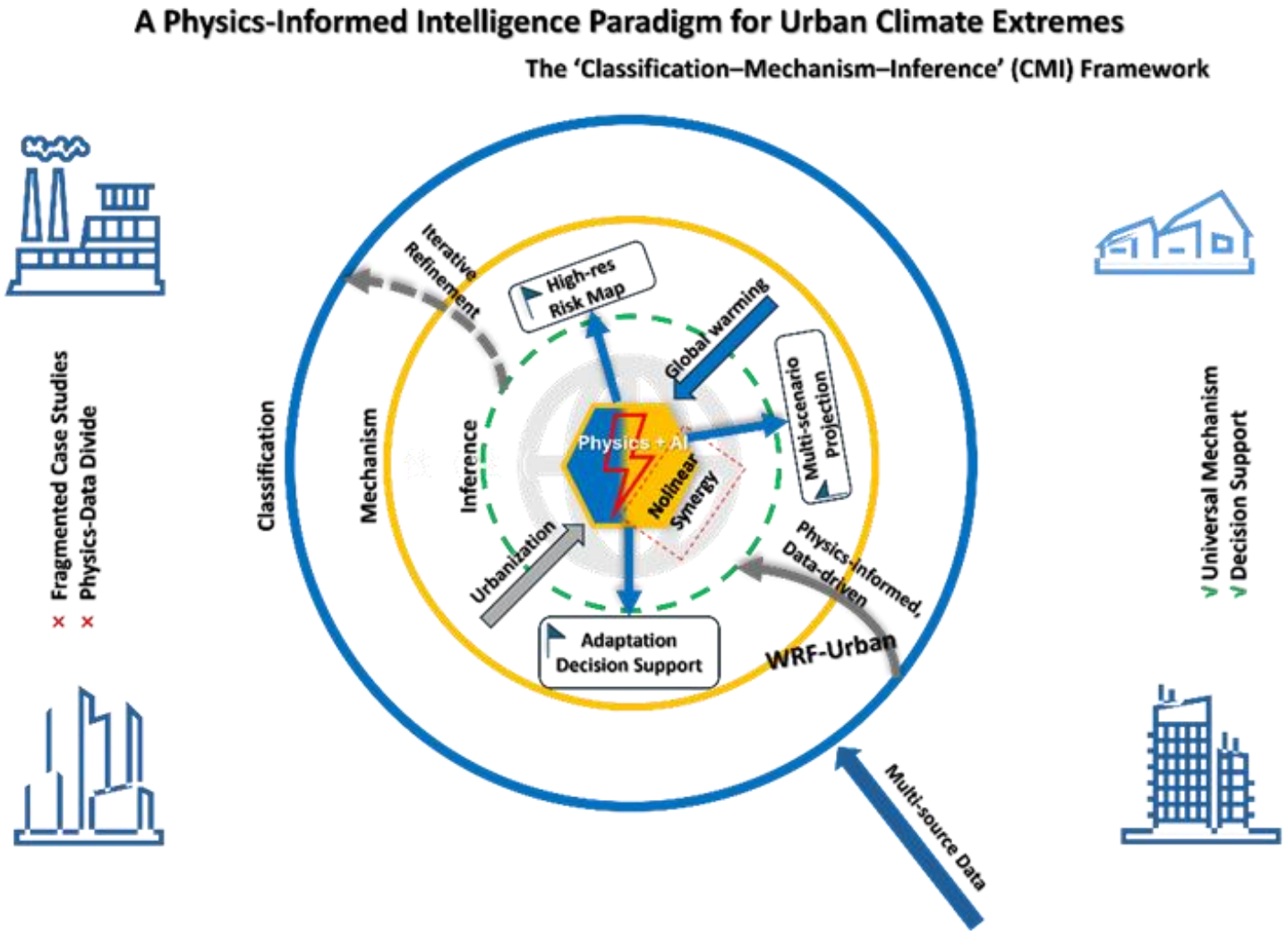

**Figure 2. Conceptual architecture of the Physics-Informed Intelligence Paradigm for Urban Climate Extremes.** The 'Classification–Mechanism–Inference' (CMI) framework integrates multi-source data and physical constraints to bridge the gap between global forcings and localized urban climate risks.

Classification: Toward a Global "Climate-Form-Function" Urban Typology. This provides the essential cognitive scaffolding for systematic comparison. Achieving it requires establishing two parallel, linkable schemas: a standardized physical taxonomy (encompassing climate, 3D morphology, and surface properties) and a distinct socio-functional analytical framework (capturing attributes like economic development and governance). The pivotal scientific advance lies in forging explicit analytical links between these schemas to trace the complete causal pathway from socio-economic drivers to physical urban signatures and, ultimately, to climatic outcomes. A foundational first step is to tame the field's persistent fragmentation in terminology and methodology, thereby establishing a unified, comparable descriptive baseline (Fleischmann et al., 2021). Now, The identification of archetypes and recurrent patterns in global urban form is now possible through unsupervised, data-driven approaches (Debray et al., 2025; Rinaldi and Ghandehari, 2026). These advances are built upon a growing suite of enabling tools: standardized morphometric indices, open-source computational software (Biljecki and Chow, 2022), and increasingly sophisticated classification-mapping techniques (Zhao et al., 2019). A robust classification framework born of these tools would do more than catalogue cities; it would enable the strategic selection of representative global samples for controlled scientific comparison. Crucially, the demonstrated ability of fine-grained schemes like Local Climate Zones to improve high-resolution physical models (Kedia et al., 2026) provides applied, empirical validation for the core scientific value of systematic classification in deepening mechanistic understanding. China's remarkably diverse urban landscape provides an ideal setting to both validate and enrich such a global classification framework.

Mechanism: Physics-Informed Machine Learning (PIML) Surrogates. PIML surrogate models represent a critical breakthrough for deciphering nonlinear synergies and transcending computational limits. The approach proceeds in two steps: first, generating high-resolution benchmark data for representative city types using physics-based models; second, training machine learning surrogates whose architectures are constrained by fundamental conservation laws (e.g., energy, momentum) (Raissi et al., 2019). This paradigm (Karniadakis et al., 2021) delivers a transformative capability—it enables the rapid generation of massive scenario ensembles at a fraction of the traditional cost while guaranteeing physical consistency. More than a computational shortcut, it provides an interpretable window into the interaction mechanisms and nonlinear thresholds governing different forcings. By doing so, it establishes a viable pathway for scalable, cross-scale urban climate simulation (Wang C. et al., 2025), directly contributing to the development of a generalizable theory across urban types.

Inference: Enabling Granular Risk Assessment and Differentiated Adaptation. This phase translates mechanistic knowledge into actionable insights for decision-makers. Rigorously validated Physics-Informed Machine Learning (PIML) surrogates empower the rapid inference of probabilistic climate risk distributions for distinct city types under varying emission and development pathways—a capability already proven in contexts like street-level flood forecasting (Bhattarai et al., 2024). This allows for the quantitative assessment of adaptation strategy effectiveness. By bridging micro-scale process understanding with macro-scale risk estimation, this mechanism-constrained inference establishes a clear causal chain from process to impact. Contemporary research underscores that adaptation must be rooted in cities as unique socio-technical-ecological systems, responsive to local contexts and developed through co-production with stakeholders, including marginalized communities, to advance equitable resilience (Lobo et al., 2023; Berner et al., 2025). Thus, this inferential capacity supplies a dynamic, quantitative evidence base for targeted, proactive adaptation planning. Fully realizing the transition from data to decision, however, requires integrating this capability within a systematic decision-support methodology.

The true value of this vast library of risk scenarios lies in its ability to be systematically channeled into actionable intelligence for concrete planning. Realizing this requires dedicated mechanisms: robust Decision Support Systems (DSS), structured evaluative frameworks, and inclusive participatory processes. Generating actionable intelligence for concrete planning is a multi-level, interdisciplinary endeavor that demands a synergistic integration of technical tools, evaluative frameworks, and governance. This integration advances along two fronts. One front addresses methodology, advocating a move beyond purely top-down Integrated Assessment Models (IAMs). This is achieved by incorporating bottom-up, participatory frameworks—such as multi-criteria decision analysis—to embed diverse societal values and strengthen the robustness of climate policy (Doukas and Nikas, 2020). The other front focuses on implementation, deploying a suite of tools scaled to different decision contexts. At the city-wide strategic level, robust Decision Support Systems (DSS) enable the comparison and ranking of development pathways (Wang and Ren, 2025). Project-level decision-making is supported by a key triad of tools: evaluative frameworks for cost-benefit analysis (e.g., for nature-based solutions; Sahay, 2024), high-resolution "digital sandbox" models for design testing (Smith et al., 2024), and participatory spatial tools for mapping benefits to inform planning (e.g., for green infrastructure; Brom et al., 2023). Ultimately, these components converge within an Environmental Decision Support System (EDSS)—a dynamic platform that synthesizes simulation, management, and participatory engagement to formulate resilient urban strategies (Rautrao et al., 2026).

## 5. Concluding Remarks

The nonlinear synergy between global warming and urbanization is a primary driver of intensifying extreme climate risks in cities, challenging sustainable development globally. Despite progress in observations, attribution, and modelling, critical gaps persist in both the systematic mechanistic understanding of these interactions and the tools needed to deeply integrate physical and data-driven approaches. To address this dual limitation, we propose an integrated Classification–Mechanism–Inference (CMI) framework. This framework establishes a foundation for systematic comparison through a global urban typology, uses Physics-Informed Machine Learning (PIML) to unravel key nonlinear processes, and leverages intelligent decision-support systems to translate mechanistic insight into actionable scenarios for risk management and adaptation. This physics-informed, data-driven loop advances the theoretical foundations of urban climate science and provides a vital evidence base for building resilient, sustainable cities worldwide. The framework can be grounded and refined using China's major urban clusters as critical test cases. Their unique confluence of intensive human activity, spatiotemporally compressed development, and a full climatic gradient forms a powerful "natural laboratory" for deciphering complex warming-urbanization dynamics. Insights and methods derived here will enable the extension of the CMI framework to other rapidly urbanizing, heterogeneous regions, offering a scalable scientific foundation for global urban climate resilience.

## Reference

Aboagye P. D., and A. Sharifi, 2024: Urban climate adaptation and mitigation action plans: A critical review. Renewable and Sustainable Energy Reviews, 189 (Part A), 113886. https://doi.org/10.1016/j.rser.2023.113886.

Alcoforado M.J. and H. Andrade, 2008: *Global Warming and the Urban Heat Island*. In Marzluff J.M. et al. (Eds.), Urban Ecology (pp. 249–262). Springer, Boston, MA. https://doi.org/10.1007/978-0-387-73412-5_14.

Berk S., M. M. Joshi, C. M. Goodess, P. Nowack, 2026: Amplified warming in tropical and subtropical cities under 2 ° C climate change. Proceedings of the National Academy of Sciences, 123(6), e2502873123. https://doi.org/10.1073/pnas.2502873123.

Berner H., S. De Gregorio Hurtado, and E. Gualini, 2025: Strategies for urban climate adaptation: Pathways towards equitable resilience in Barcelona and Berlin. Cities, 161, 105836. https://doi.org/10.1016/j.cities.2025.105836.

Bhattarai Y., S. Bista, R. Talchabhadel, S. Duwal, and S. Sharma, 2024: Rapid prediction of urban flooding at street-scale using physics-informed machine learning-based surrogate modeling. Total Environment Advances, 12, 200116. https://doi.org/10.1016/j.teadva.2024.200116.

Bi K., L. Xie, H. Zhang, et al., 2023: Accurate medium-range global weather forecasting with 3D neural networks. Nature, 619(7970), 533 – 538. https://doi.org/10.1038/s41586-023-06185-3.

Biljecki F. and Y.S. Chow, 2022: Global Building Morphology Indicators. Computers, Environment and Urban Systems, 95, 101809. https://doi.org/10.1016/j.compenvurbsys.2022.101809.

Bobde V., K. Ayegbusi, A.A. Akinsanola, et al., 2025: Anthropogenic warming is accelerating recent heatwaves in Africa. Commun Earth Environ, 6, 578. https://doi.org/10.1038/s43247-025-02578-6.

Bounoua L., et al., 2015: Impact of urbanization on US surface climate. Environ. Res. Lett., 10, doi:10.1088/1748-9326/10/8/084010.

Bracco, A., J. Brajard, H. A. Dijkstra, et al., 2025: Machine learning for the physics of climate. Nat Rev Phys, 7, 6 – 20. https://doi.org/10.1038/s42254-024-00776-3.

Brom P., K. Engemann, C. Breed, M. Pasgaard, T. Onaolapo, J.-C. Svenning, 2023: A Decision Support Tool for Green Infrastructure Planning in the Face of Rapid Urbanization. Land, 12(2), 415. https://doi.org/10.3390/land12020415.

Camps-Valls G., MÁ. Fernάndez-Torres, KH. Cohrs, et al., 2025: Artificial intelligence for modeling and understanding extreme weather and climate events. Nat Commun, 16, 1919. https://doi.org/10.1038/s41467-025-56573-8.

Chen K., J. Boomsma, and H.A. Holmes, 2023: A multiscale analysis of heatwaves and urban heat islands in the western U.S. during the summer of 2021. Sci Rep, 13, 9570. https://doi.org/10.1038/s41598-023-35621-7.

Chen W., J. Piao, S. Chen, L. Wang, W. Zhao, Z. Wang, and Q. Wang, 2025: Multi-scale variations and future projections of dry-wet conditions over the monsoon transitional zone in East Asia: A review. Fundamental Research, 5(4), 1597 – 1606. https://doi.org/10.1016/j.fmre.2024.01.023.

Cohen-Shacham E., G. Walters, C. Janzen, and S. Maginnis (Eds.), 2016: Nature-based solutions to address global societal challenges. Gland, Switzerland: International Union for Conservation of Nature (IUCN). https://doi.org/10.2305/IUCN.CH.2016.13.en.

Collier S.J. and Y. Nidam Kirsht, 2026: Disorderly urban adaptation to climate change: qualitative modeling of finance, land use, and the built environment. npj Urban Sustain, 6, 13. https://doi.org/10.1038/s42949-025-00318-5.

Cook E.M., Y. Kim, N.B. Grimm, et al., 2025: Nature-based solutions for urban sustainability. Proceedings of the National Academy of Sciences, 122(29), e2315909122. https://doi.org/10.1073/pnas.2315909122.

Creutzig F., T. McPhearson, R. Bardhan, et al., 2025: Bridging the scale between the local particular and the global universal in climate change assessments of cities. Nat Cities, 2, 369 – 378. https://doi.org/10.1038/s44284-025-00226-w.

de Bono A., G. Giuliani, S. Kluser, P. Peduzzi, 2004: Impacts of summer 2003 heat wave in Europe. Environment Alert Bulletin. University of Geneva.

Debray H., M. Gassilloud, R. Lemoine-Rodríguez, M. Wurm, X. Zhu, and H. Taubenböck, 2025: Universal patterns of intra-urban morphology: Defining a global typology of the urban fabric using unsupervised clustering. International Journal of Applied Earth Observation and Geoinformation, 141, 104610. https://doi.org/10.1016/j.jag.2025.104610.

Diffenbaugh N. S., D. Singh, J. S. Mankin, et al., 2017: Quantifying the influence of global warming on unprecedented extreme climate events. Proceedings of the National Academy of Sciences, 114(19), 4881-4886.

Ding X., Y. Fan, Y. Zhao, et al., 2026: Asymmetric global urban cooling potential demands accelerated and context-specific actions. Nat Commun. https://doi.org/10.1038/s41467-026-70662-2.

Doukas H. and A. Nikas, 2020: Decision support models in climate policy. European Journal of Operational Research, 280(1), 1 – 24. https://doi.org/10.1016/j.ejor.2019.01.017.

Du H., W. Zhan, Z. Liu, E.S. Krayenhoff, T.C. Chakraborty, L. Zhao, L. Jiang, P. Dong, L. Li, F. Huang, S. Wang, and Y. Xu, 2023: Global mapping of urban thermal anisotropy reveals substantial potential biases for remotely sensed urban climates. Science Bulletin, 68(16), 1809 – 1818. https://doi.org/10.1016/j.scib.2023.06.032.

Elmqvist T., et al. (eds.), 2013: Urbanization, Biodiversity and Ecosystem Services: Challenges and Opportunities: A Global Assessment. Springer Dordrecht Heidelberg New York London, ISBN 978-94-007-7088-1 (eBook), doi: 10.1007/978-94-007-7088-1.

Eyring, V., P. Gentine, G. Camps-Valls, et al., 2024a: AI-empowered next-generation multiscale climate modelling for mitigation and adaptation. Nat. Geosci., 17, 963 – 971. https://doi.org/10.1038/s41561-024-01527-w.

Eyring, V., W. D. Collins, P. Gentine, et al., 2024b: Pushing the frontiers in climate modelling and analysis with machine learning. Nat. Clim. Chang., 14, 916 – 928. https://doi.org/10.1038/s41558-024-02095-y.

Fleischmann M., O. Romice, and S. Porta, 2021: Measuring urban form: Overcoming terminological inconsistencies for a quantitative and comprehensive morphologic analysis of cities. Environment and Planning B: Urban Analytics and City Science, 48(8), 2133 – 2150. https://doi.org/10.1177/2399808320910444.

Founda D. and M. Santamouris, 2017: Synergies between Urban Heat Island and Heat Waves in Athens (Greece), during an extremely hot summer (2012). Sci Rep, 7, 10973. https://doi.org/10.1038/s41598-017-11407-6.

Gál T., N. Alinasab, H.A. Abdulhaq, and N. Skarbit, 2026: A Structured Literature Review of the Application of Local Climate Zones (LCZ) in Urban Climate Modelling. Earth, 7(1), 3. https://doi.org/10.3390/earth7010003.

Grimmond S., 2007: Urbanization and global environmental change: local effects of urban warming. Cities and global environmental change, 83-88.

Haase, D., N. Larondelle, E. Andersson, M. Artmann, S. Borgström, J. Breuste, E. GomezBaggethun, A. Gren, Z. Hamstead, R. Hansen, N. Kabisch, P. Kremer, J. Langemeyer, E. Lorance Rall, T. McPhearson, S. Pauleit, S. Qureshi, N. Schwarz, A. Voigt, D. Wurster, and

T. Elmqvist, 2014: A quantitative review of urban ecosystem services assessment: concepts, models and implementation. AMBIO, 43(4), 413-433.

Han T., S. Guo, F. Ling, et al., 2024: FengWu-GHR: Learning the Kilometer-scale Medium-range Global Weather Forecasting. arXiv preprint, arXiv:2402.00059. https://arxiv.org/abs/2402.00059.

He C., M. Demuzere, A. Zonato, and A. Martilli, 2023: WRF-urban model and global LCZ data implementation. Presentation at the 2023 WUDAPT Workshop, NCAR. Retrieved from https://ral.ucar.edu/sites/default/files/docs//wudaptworkshoplczwrfurbancenlinhe20230113.pdf.

He W. H., L. Q. Zhang and C. Yuan, 2022: Future air temperature projection in high-density tropical cities based on global climate change and urbanization-a study in Singapore. Urban Climate, 42, 101115, https://doi.org/10.1016/j.uclim.2022.101115.

Hobbie S. E. and N. B. Grimm, 2020: Nature-based approaches to managing climate change impacts in cities. Phil. Trans. R. Soc. B, 375, 20190124, http://dx.doi.org/10.1098/rstb.2019.0124.

Huang J., Z. Huang, and W. Liu, 2025: Combining the WRF model and LCZ scheme to assess spatiotemporal variations of thermal comfort in Shenzhen’s built-up areas. Sustainable Cities and Society, 122, 106252. https://doi.org/10.1016/j.scs.2025.106252.

Huang R. H. and Z.C. Du, 2010: Evolution characteristics and trends of drought and flood climatic disasters in China under global warming (in Chinese). Chinese Journal of Nature, 32(4), 187-195.

Huang S. Z., X. Zhang, L. Yang, N. C. Chen, W.-H. Nam, D. Niyogi, 2022: Urbanization-induced drought modification: Example over the Yangtze River Basin, China. Urban Climate, 44, 101231, https://doi.org/10.1016/j.uclim.2022,101231.

Huang X., L. Hao, G. Sun, Z.-L. Yang, W. Li, D. Chen, 2022. Urbanization aggravates effects of global warming on local atmospheric drying. Geophysical Research Letters, 49, e2021GL095709. https://doi.org/10.1029/2021GL095709.

Igun E., X. Xu, Z. Shi, and G. Jia, 2023: Enhanced nighttime heatwaves over African urban clusters. Environmental Research Letters, 18(1), 014001. https://doi.org/10.1088/1748-9326/aca920.

IPCC, 2021: Summary for Policymakers. In: Climate Change 2021: The Physical Science Basis. Contribution of Working Group I to the Sixth Assessment Report of the Intergovernmental Panel on Climate Change, [Masson-Delmotte, V., P. Zhai, A. Pirani, S.L. Connors, C. Péan, S. Berger, N. Caud, Y. Chen, L. Goldfarb, M.I. Gomis, M. Huang, K. Leitzell, E. Lonnoy, J.B.R. Matthews, T.K. Maycock, T. Waterfield, O. Yelekçi, R. Yu, and B. Zhou (eds.)]. Cambridge University Press, Cambridge, United Kingdom and New York, NY, USA, pp. 3–32, doi:10.1017/9781009157896.001.

IPCC, 2022: Climate Change 2022: Impacts, Adaptation and Vulnerability. Contribution of Working Group II to the Sixth Assessment Report of the Intergovernmental Panel on Climate Change, [H.-O. Pörtner, D.C. Roberts, M. Tignor, E.S. Poloczanska, K. Mintenbeck, A. Alegría, M. Craig, S. Langsdorf, S. Löschke, V. Möller, A. Okem, B. Rama (eds.)]. Cambridge University Press, Cambridge, UK and New York, NY, USA, 3056 pp. doi:10.1017/9781009325844.

International Union for Conservation of Nature (IUCN), 2020: IUCN Global Standard for Nature-based Solutions: A user-friendly framework for the verification, design and scaling up of NbS (1st ed.). IUCN. https://doi.org/10.2305/IUCN.CH.2020.08.en.

Joshi P., T.-S. Lin, C. He, and K. Lamer, 2025: Urban weather modeling using WRF: linking physical assumptions, code implementation, and observational needs, Geosci. Model Dev., 18, 7869 – 7890, https://doi.org/10.5194/gmd-18-7869-2025.

Kalnay E. and M Cai, 2003: Impact of urbanization and land-use change on climate. Nature, 423, 528-531.

Karniadakis G.E., I.G. Kevrekidis, L. Lu, et al., 2021: Physics-informed machine learning. Nat Rev Phys, 3, 422 – 440. https://doi.org/10.1038/s42254-021-00314-5.

Kedia S., S. Islam, P. Sinha, S. Gavhale, and M. Khare, 2026: Impact of realistic urban morphology on hyper-local scale rainfall prediction using WRF. Urban Climate, 67, 102871. https://doi.org/10.1016/j.uclim.2026.102871.

Khan A., K. Vasilakopoulou, and M. Santamouris, 2025: Exploring the potential impacts of anthropogenic heating on urban climate during heatwaves. Sci Rep 15, 3908. https://doi.org/10.1038/s41598-024-83918-y.

Kim B. and T. Kim, 2025: AI in extreme weather events prediction and response: a systematic topic-model review (2015 – 2024). Frontiers in Environmental Science, 13, 1659344. https://doi.org/10.3389/fenvs.2025.1659344.

Kochkov D., J. Yuval, I. Langmore, et al., 2024: Neural general circulation models for weather and climate. Nature, 632, 1060 – 1066. https://doi.org/10.1038/s41586-024-07744-y.

Kong D., X. Gu, J. Li, G. Ren, and J. Liu, 2020: Contributions of global warming and urbanization to the intensification of human-perceived heatwaves over China. Journal of Geophysical Research: Atmospheres, 125, https://doi.org/10.1029/2019JD032175.

Lam R., A. Sanchez-Gonzalez, M. Willson, et al., 2023: Learning skillful medium-range global weather forecasting. Science, 382(6677), 1416 – 1421. https://doi.org/10.1126/science.adi2336.

Lempert R. J., 2021: Measuring global climate risk. Nat. Clim. Chang. 11, 805 – 806. https://doi.org/10.1038/s41558-021-01165-9.

Li L., W. Zhan, L. Hu, T.C. Chakraborty, Z. Wang, P. Fu, D. Wang, W. Liao, F. Huang, H. Fu, J. Li, Z. Liu, H. Du, and S. Wang, 2023: Divergent urbanization-induced impacts on global surface urban heat island trends since 1980s. Remote Sensing of Environment, 295, 113650. https://doi.org/10.1016/j.rse.2023.113650.

Li M., Y. Yao, I. Simmonds, D. Luo, L. Zhong, and X. Chen, 2020: Collaborative impact of the NAO and atmospheric blocking on European heatwaves, with a focus on the hot summer of 2018. Environmental Research Letters, 15(11), 114003. https://doi.org/10.1088/1748-9326/aba6ad.

Li X., J. Hang, Z. Mo, Q. Wang, 2026: Impact of adjacent city development on the urban heat island intensity and thermal stress: A WRF-BEP/BEM and LCZ-integrated assessment. Building and Environment, 289, 114024. https://doi.org/10.1016/j.buildenv.2025.114024.

Liao S., W. Pan, L. Wen, et al., 2025: Temperature-related hospitalization burden under climate change. Nature, 644, 960 – 968. https://doi.org/10.1038/s41586-025-09352-w.

Liao W., D. Wang, X. Liu, et al., 2017: Estimated influence of urbanization on surface warming in Eastern China using time-varying land use data. Int. J. Climatol., 37, 3197–3208. https://doi.org/10.1002/joc.4908.

Liao W., L. Wang, X. Liu, et al., 2025: Standardized heat islands and persistence drive modeled urban heat events. Nat Cities, 2, 857 – 864. https://doi.org/10.1038/s44284-025-00290-2.

Liu R., Z. Han, J. Li, J. Li, L. Liang, and Y. Wu, 2023: The impacts of urban anthropogenic heat and surface albedo change on boundary layer meteorology and air pollutants in the Beijing-Tianjin-Hebei region. Urban Climate, 47, 101358. https://doi.org/10.1016/j.uclim.2022.101358.

Lobo J., R.M. Aggarwal, M. Alberti, et al., 2023: Integration of urban science and urban climate adaptation research: opportunities to advance climate action. npj Urban Sustain, 3, 32. https://doi.org/10.1038/s42949-023-00113-0.

Lopez-Gomez I., A. McGovern, S. Agrawal, and J. Hickey, 2023: Global Extreme Heat Forecasting Using Neural Weather Models. Artificial Intelligence for the Earth Systems, 2(1), e2200035. https://doi.org/10.1175/AIES-D-22-0035.1

Luo J.-J., J. Xia, B. Pan, et al., 2026: AI for atmosphere-ocean sciences: advancements, challenges, and ways forward. National Science Review, nwag063. https://doi.org/10.1093/nsr/nwag063.

Luo M. et al., 2024: Anthropogenic forcing has increased the risk of longer-traveling and slower-moving large contiguous heatwaves. Sci. Adv., 10(13), eadl1598. https://doi.org/10.1126/sciadv.adl1598.

Luo, M. and N.-C. Lau, 2021: Increasing human-perceived heat stress risks exacerbated by urbanization in China: A comparative study based on multiple metrics. Earth's Future, 9, e2020EF001848. https://doi. org/10.1029/2020EF001848.

Mancheno G., J. Castro-Molinare, and H. Jorquera, 2024: Predictive modeling the effect of Local Climate Zones (LCZ) on the urban meteorology in a tropical andean area. Model. Earth Syst. Environ., 10, 6133 – 6153. https://doi.org/10.1007/s40808-024-02121-0.

Meehl G. A. and C. Tebaldi, 2004: More intense, more frequent, and longer lasting heat waves in the 21st century. Science, 305(5686), 994 – 997, https://doi.org/10.1126/science.1098704.

Miao S., W. Jiang, P. Liang, et al., 2020: Advances in urban meteorology research (in Chinese). Acta Meteorologica Sinica, 78(3), 477-499. https://doi.org/10.11676/qxxb2020.025.

Ministry of Ecology and Environment of the People's Republic of China, et al., 2022:.National Climate Change Adaptation Strategy 2035 (in Chinese). Retrieved from https://www.mee.gov.cn/xxgk2018/xxgk/xxgk03/202206/W020220613636562919192.pdf.

Mora C., B. Dousset, I. R. Caldwell, F. E. Powell, R. C. Geronimo, C. R. Bielecki, et al., 2017: Global risk of deadly heat. Nature Climate Change, 7(7), 501 – 506. https://doi.org/10.1038/nclimate3322.

Nath R., D. Nath, and W. Chen, 2024: Projected changes in extreme hot summer events in Asian monsoon regions. npj Clim Atmos Sci, 7, 188. https://doi.org/10.1038/s41612-024-00734-x.

Nath S. and T. Palmer, 2026: Can AI models reliably forecast extreme weather events? Nature, 651, 583-584. https://doi.org/10.1038/d41586-026-00842-z.

Obe O.B., T.E. Morakinyo, and G. Mills, 2024: An assessment of WRF-urban schemes in simulating local meteorology for heat stress analysis in a tropical sub-Saharan African city, Lagos, Nigeria. Int J Biometeorol, 68, 811–828. https://doi.org/10.1007/s00484-024-02627-3.

Oke T. R., 1973: City size and the urban heat island. Atmospheric Environment (1967), 7(8), 769–779, https://doi.org/10.1016/0004-6981(73)90140-6.

Park B.-J., Y.-H. Kim, S.-K. Min, M.-K, Kim, Y. Choi, K.-O. Boo, and S. Shim, 2017: Long-term warming trends in Korea and contribution of urbanization: An updated assessment. Journal of Geophysical Research: Atmospheres, 122, 10,637–10,654, https://doi.org/10.1002/2017JD027167.

Pathak J., S. Subramanian, P. Harrington, et al., 2022: FourCastNet: A Global Data-driven High-resolution Weather Model using Adaptive Fourier Neural Operators. https://doi.org/10.48550/arXiv.2202.11214.

Pfautsch S., A. Wujeska-Klause, and J.R. Walters, 2025: Spatiotemporal variation of intra-urban heat and heatwaves across Greater Sydney, Australia. Weather and Climate Extremes, 47, 100741, https://doi.org/10.1016/j.wace.2024.100741.

Possega M., L. Aragão, P. Ruggieri, M.A. Santo, and S. Di Sabatino, 2022: Observational evidence of intensified nocturnal urban heat island during heatwaves in European cities. Environmental Research Letters, 17(12), 124013. https://doi.org/10.1088/1748-9326/aca3ba.

Raissi M., P. Perdikaris, and G.E. Karniadakis, 2019: Physics-informed neural networks: A deep learning framework for solving forward and inverse problems involving nonlinear partial differential equations. Journal of Computational Physics, 378, 686–707. https://doi.org/10.1016/j.jcp.2018.10.045.

Rautrao R.R., A. Gawande, A. Kumar, S. Jadhav, A.K. Sikhakolli, 2026: Development of Environmental Decision Support System for Climate Change Resilience in Urban Planning: A Managerial Perspective. In: Srivastav, A.L., Dubey, A.K., Kumar, A., Ouaissa, M. (eds)

Climate Resilience: Impact of Quantum Computing and Artificial Intelligence on Urban Planning. Studies in Systems, Decision and Control, vol 637. Springer, Cham. https://doi.org/10.1007/978-3-032-06791-3_18.

Raymond C., R.M. Horton, J. Zscheischler, et al., 2020: Understanding and managing connected extreme events. Nat. Clim. Chang., 10, 611 – 621. https://doi.org/10.1038/s41558-020-0790-4.

Rinaldi V. and M. Ghandehari, 2026: Characterizing urban form archetypes through landscape and morphology parameterization and dimensionality reduction. International Journal of Digital Earth, 19(1), 2616955. https://doi.org/10.1080/17538947.2026.2616955.

Sahay S., 2024: Nature-based solutions as urban adaptation to climate risk: Framework for economic evaluation as decision support tool. Sustainable Cities and Society, 118, 106037. https://doi.org/10.1016/j.scs.2024.106037.

Sarangi C., Y. Qian, J. Li, L. R. Leung, T. C. Chakraborty, and Y. Liu, 2021: Urbanization amplifies nighttime heat stress on warmer days over the US. Geophysical Research Letters, 48, e2021GL095678, https://doi.org/10.1029/2021GL095678.

Sethi S.S. and V. Vinoj, 2024: Urbanization and regional climate change-linked warming of Indian cities. Nat Cities, 1, 402 – 405. https://doi.org/10.1038/s44284-024-00074-0.

Smith P., E. Blanco, P. Sarricolea, O. Peralta, and F. Thomas, 2024: Urban climate simulation model to support climate-sensitive planning decision making at local scale. Journal of Urban Management. 14(1), 279-292. https://doi.org/10.1016/j.jum.2024.11.003.

Stewart I.D. and T.R. Oke, 2012: Local Climate Zones for urban temperature studies. Bulletin of the American Meteorological Society, 93(12), 1879-1900.

Trenberth K. E., 2008: The impact of climate change and variability on heavy precipitation, floods and droughts. Encyclopedia of hydrological sciences, M. H. Anderson, Ed., J. Wiley and Sons, doi: 10.1002/0470848944.hsa211.

UNEA (United Nations Environment Assembly), 2022: Resolution on Nature-based Solutions for supporting sustainable development (UNEP/EA.5/Res.5). Nairobi, Kenya: United Nations Environment Programme. Retrieved from https://digitallibrary.un.org/record/3999268?v=pdf.

UNFCCC (United Nations Framework Convention on Climate Change), 2015: Paris Agreement. Bonn, Germany: UNFCCC Secretariat. Retrieved from https://unfccc.int/sites/default/files/english_paris_agreement.pdf.

Vautard R., J. Cattiaux, T. Happé, et al., 2023: Heat extremes in Western Europe increasing faster than simulated due to atmospheric circulation trends. Nat Commun, 14, 6803. https://doi.org/10.1038/s41467-023-42143-3.

Wang C., Y. Zhao, Q. Li, Z.-H. Wang, and J. Fan, 2025: Ultrafine-resolution Urban Climate modeling: Resolving processes across scales. Journal of Advances in Modeling Earth Systems, 17, e2025MS005053. https://doi.org/10.1029/2025MS005053.

Wang J., Y. Chen, W. Liao, et al., 2021: Anthropogenic emissions and urbanization increase risk of compound hot extremes in cities. Nat. Clim. Chang., 11, 1084 – 1089. https://doi.org/10.1038/s41558-021-01196-2.

Wang S., W. Zhan, B. Zhou, et al., 2025: Dual impact of global urban overheating on mortality. Nat. Clim. Chang., 15, 497 – 504. https://doi.org/10.1038/s41558-025-02303-3.

Wang Y., Y. Ren, L. Song, and Y. Xiang, 2021: Responses of extreme high temperatures to urbanization in the Beijing – Tianjin – Hebei urban agglomeration in the context of a changing climate. Meteorological Applications, 28(5), e2024, https://doi.org/10.1002/met.2024.

Wang Z. and F. Ren, 2025: Developing a decision support system for sustainable urban planning using machine learning-based scenario modeling. Sci Rep, 15, 13210. https://doi.org/10.1038/s41598-025-90057-5.

Xian G. Z., 2023: Characterizing Urban Heat Islands Across 50 Major Cities in the United States (U.S. Geological Survey Fact Sheet 2023 – 3048). U.S. Geological Survey. https://doi.org/10.3133/fs20233048.

Xiong J., Y. Yang, L. Yang, et al., 2025: Asymmetric shifts in precipitation due to urbanization across global cities. Nat Commun, 16, 5802. https://doi.org/10.1038/s41467-025-61053-0.

Xuan D., M. A. Hsieh, L. S. Pongeluppe, et al., 2026: Climate extremes and urbanization drive flood tipping points at the city – river interface. npj Natural Hazards, 3(1), 20. https://doi.org/10.1038/s44304-026-00186-8.

Yan Y., C. Dong, Z. Liu, et al., 2026: Disparities in urban vegetation degradation and heat exposure during drought periods in U.S. cities. npj Urban Sustain, 6, 14. https://doi.org/10.1038/s42949-025-00319-4.

Yang L., Y. Yang, Y. Shen, et al., 2024: Urban development pattern’s influence on extreme rainfall occurrences. Nat Commun, 15, 3997. https://doi.org/10.1038/s41467-024-48533-5.

Yang X., Y. Hou, and B. Chen, 2011: Observed surface warming induced by urbanization in east China, J. Geophys. Res., 116, D14113, doi:10.1029/2010JD015452.

Yenneti K., L. Ding, D. Prasad, G. Ulpiani, R. Paolini, S. Haddad, M. Santamouris, 2020: Urban Overheating and Cooling Potential in Australia: An Evidence-Based Review. Climate, 8(11):126. https://doi.org/10.3390/cli8110126.

Yu W., J. Yang, Y. Zhou, et al., 2026: Global dataset on heat wave exposure due to the urban heat island effect. Sci Data. Advance online publication https://doi.org/10.1038/s41597-026-06877-1.

Yuan J. C., Q. Y. Lin, S. S. Chen, H. Zhao, X. Xie, Z. Y. Cai, J. T. Zhang, T. T. Cheng, M. Hua, R. H. Zhang, 2022: Influence of global warming and urbanization on regional climate of Megacity: A case study of Chengdu, China. Urban Climate, 44, https://doi.org/10.1016/j.uclim.2022.101227.

Zhan W., L. Li, T.C. Chakraborty, L. Hu, D. Wang, W. Liao, et al., 2025: Recent widespread deceleration of global surface urban heat islands unveiled by satellites. Geophysical Research Letters, 52, e2024GL112711. https://doi.org/10.1029/2024GL112711.

Zhang X., T. Zhou, W. Zhang, et al., 2023: Increased impact of heat domes on 2021-like heat extremes in North America under global warming. Nat Commun, 14, 1690. https://doi.org/10.1038/s41467-023-37309-y.

Zhang Z., E. Fischer, J. Zscheischler, and S. Engelke, 2025: Numerical models outperform AI weather forecasts of record-breaking extremes. arXiv preprint, arXiv:2508.15724.

Zhao Z., J. Ma, F. Zhong, and J. Cao, 2019: Self-Training Classification Framework with Spatial-Contextual Information for Local Climate Zones. Remote Sensing, 11(23), 2828. https://doi.org/10.3390/rs11232828.

Zhou D., J. Xiao, S. Frolking, L. Zhang, and G. Zhou, 2022: Urbanization contributes little to global warming but substantially intensifies local and regional land surface warming. Earth's Future, 10, e2021EF002401, https://doi.org/10.1029/2021EF002401.